\RequirePackage{fix-cm}
\documentclass[smallextended]{svjour3}       % onecolumn (second format)
\smartqed  % flush right qed marks, e.g. at end of proof
\usepackage{amsmath}
\usepackage{booktabs} % For pretty tables
\usepackage{caption} % For caption spacing
\usepackage{subcaption} % For sub-figures
\usepackage{graphicx}
\usepackage{pgfplots}
\usepackage[all]{nowidow}
\usepackage[utf8]{inputenc}
\usepackage{tikz}
\usetikzlibrary{er,positioning}
\usepackage{multicol}
\usepackage{graphicx,float}
\graphicspath{ {./} }
\usepackage{hyperref}
\usepackage[inline]{enumitem}
\usepackage[ruled,vlined]{algorithm2e}

\def\makeheadbox{\relax}

\begin{document}

\title{A least squares support vector regression for anisotropic diffusion filtering}

\titlerunning{A LS-SVR for anisotropic diffusion filtering}        % if too long for running head

\author{Arsham Gholamzadeh Khoee  \and
Kimia Mohammadi Mohammadi  \and
Mostafa Jani  \and
Kourosh Parand }

\authorrunning{A. G. Khoee et al.} % if too long for running head

\institute{A. G. Khoee \at
              Department of Computer Science, School of Mathematics, Statistics, and Computer Science, University of Tehran, Tehran, Iran \\
              \email{arsham.khoee@ut.ac.ir}
           \and
           K. M. Mohammadi \at
              Department of Computer Sciences, Shahid Beheshti University, G.C., Tehran, Iran \\
              \email{ki.mohammadimohammad@mail.sbu.ac.ir}
            \and
           M. Jani \at
              Department of Computer Sciences, Shahid Beheshti University, G.C., Tehran, Iran \\
              \email{m\_jani@sbu.ac.ir}
             \and
           K. Parand \at
              Department of Computer Sciences, Shahid Beheshti University, G.C., Tehran, Iran \\
              Department of Cognitive Modelling, Institute for Cognitive and Brain Sciences, Shahid Beheshti University, G.C, Tehran, Iran \\
              School of Computer Science, Institute for Research in Fundamental Sciences (IPM), Tehran, Iran \\
              \email{k\_parand@sbu.ac.ir}
}


\maketitle

\begin{abstract}
Anisotropic diffusion filtering for signal smoothing as a low-pass filter has the advantage of the edge-preserving, i.e., it does not affect the edges that contain more critical data than the other parts of the signal. In this paper, we present a numerical algorithm based on least squares support vector regression by using Legendre orthogonal kernel with the discretization of the nonlinear diffusion problem in time by the Crank-Nicolson method. This method transforms the signal smoothing process into solving an optimization problem that can be solved by efficient numerical algorithms. In the final analysis, we have reported some numerical experiments to show the effectiveness of the proposed machine learning based approach for signal smoothing.

\end{abstract}

\section{Introduction}
\label{sec:intro}
Signal smoothing is generally about diminishing the high frequencies of a signal in order to immediately discover the trend hidden in the signal and extract some important features.  It also makes the signal utilizable for further analysis. For instance, the derivative of the signal may be required while the input signal is not differentiable. In this situation, smoothing facilitates further analysis.

Among various signal smoothing techniques, the anisotropic diffusion equation has found great importance, due to edge-preserving, denoising even the smallest features of MR images, etc. in signal processing \cite{abdallah2016adaptive}, as a fascinating technique in adaptive smoothing approaches. Adaptive smoothing methods are in contrast to the global smoothing methods, in which they do not apply over smooth sharp edges and other valuable features of a signal \cite{meyer2014pattern}. Thus, anisotropic diffusion filtering has attracted considerable attention from many researchers. At first, Witkin introduced the relation between the Gaussian function and heat equation \cite{witkin1987scale}. Later on, Perona and Malik found a smoothing method based on a special nonlinear diffusion equation \cite{perona1990scale}. Also, a time-fractional diffusion model is used for signal smoothing \cite{li2018time}. Moreover, the diffusion filtering has been used for graph signals with an application in recommendation systems \cite{ma2016diffusion}. It has also found applications for removing noise in medical magnetic resonance images \cite{lysaker2003noise}.

Here, we present some preliminaries required in the rest of the paper.

\subsection{The signal and the noise}
\label{subsec:signal-noise}
In signal processing, noise is a destructive thing in analyzing, modifying, and synthesizing signals. There are several techniques for omitting unwanted components from a signal, but it is essential not to lose vital data after the smoothing process and avoid reshaping. However, many of the existing techniques fail to have this feature. We use anisotropic diffusion filtering, which is a well-known technique for signal smoothing that performs better than other techniques and preserves edges of the input signal.

The quality of a signal is measured by the signal-to-noise ratio (SNR) that represents the ratio of the signal power to the noise power, normally expressed in decibels as:

\begin{equation}
    \textrm{SNR} = 10 \log \frac{\sum_{n}^{} s^2 (n)}{\sum_{n}^{} \| f(n) - s(n) \|  ^2}, 
\end{equation}
where $s(n)$ is the smoothed signal and $f(n)$ is the noisy signal.

As a rule, a larger value of SNR means the power of the signal dominates the power of noise. Note that the difference of the measures in decibel is equal to the logarithm of the division.

We demonstrate the isotropic and anisotropic diffusion equations in the next section.
\subsection{Nonlinear diffusion equation and application in signal smoothing}
\label{subsec:diffusion-equ}
Here, we describe the isotropic and anisotropic diffusion equation. The later, also called the Perona–Malik equation,  provides a technique to reduce noises from a signal without dissolving edges and other significant features.

The isotropic diffusion equation, as a heat equation, describes the temperature changes in a material as time goes by. Isotropic diffusion equation, in signal processing is equivalent to the heat equation as a partial differential equation (PDE), given as:
\begin{equation}
    \frac{\partial u}{\partial t} = k \nabla^2 u,
\end{equation}
where $u(x,t)$ represents the time evolution of a physical state such as the temperature with the initial condition $u(x,0)$ as the input signal (with noise added), and $k$  controls the speed and spatial scale of changing the unit of measure in time. In the heat equation, the coefficient depends on the thermal conductivity of the material.

Perona \& Malik introduced anisotropic diffusion in which the heat flows unevenly and force the diffusion process to contiguous homogeneous regions, but not cross the region boundaries \cite{parand2017numerical}.
The formula is given as:
\begin{equation}
    \frac{\partial u}{\partial t} = \nabla c \cdot \nabla u + c\left(x,t\right)\Delta u ,
\end{equation}
where $c$ refers to Perona-Malik coefficient which controls the rate of diffusion and has been chosen as a function of the signal gradient so as to preserve edges in the signal, Perona \& Malik suggested two functions as the diffusion coefficient:
\begin{equation}
\label{eq:c1}
    c\left(\|\nabla u\|\right) = e^{-\left(\|\nabla u\|/K\right)^2} , 
\end{equation}
    and
\begin{equation}
\label{eq:c2}
     c\left(\|\nabla u\|\right) = \frac{1}{1 + \left(\frac{\|\nabla u\|}{K}\right)^2} , 
\end{equation}
in which the constant $K$ adjusts the sensitivity to edges. Intuitively, a large value of $K$ leads back into an isotropic-like solution \cite{perona1990scale}.

Obviously, with a constant diffusion coefficient, the anisotropic diffusion is reduced to the heat equation, which is referred to as isotropic diffusion, where the heat flows evenly at all points of the material.

The evolution of the noisy signal in the anisotropic diffusion process leads to a smoothed signal.

Now, we discuss the classical methods concerning signal smoothing.

\subsection{Classic signal smoothing methods}
\label{subsec:classic-methods}
Most of the classical smoothing algorithms are based on averaging, i.e. any point of the signal is replaced by a weighted average of a subset of adjacent points so providing a smooth result.

The moving average algorithm is a fundamental smoothing algorithm, converting an array of noisy data into an array of smoothed data. Assuming ${\left(y_k \right)_s}$  as a set of points of smoothed data set and $n$ is the determined length of the coefficient array, calculating points of the smoothed signal is done as follows:
\begin{equation}
    \left(y_k \right)_s = \sum_{i = -n}^{n} \frac{y_{k + i}}{2n + 1} .
\end{equation}

In 1964, Savitzky and Golay \cite{schafer2011savitzky} published a table of weighting coefficients, derived by polynomial regression, in order to enhance edge-preserving. Their work leads to another smoothing algorithm using a low degree polynomial. Assuming ${\left(y_k \right)_s}$  as a set of points of smoothed data set and $A_i$ as a set of particular weighting coefficients. The equation will be as follow:
\begin{equation}
    \left(y_k \right)_s = \frac{\sum_{i = -n}^{n} A_i y_{k + i}}{\sum_{i = -n}^{n} A_i} .
\end{equation}
Note that this technique is implemented on a nonsmooth-cure after a sampling by function evaluation on some points.

Another distinguished method is the Fourier Transform (FT) which manipulates specific frequency components \cite{hieftje1973digital}. Any non-sinusoidal periodic signal can be approximated to many discrete sinusoidal frequency elements. In fact, by completely removing the high frequency elements, the smoothed signal will be achieved. Despite some advantages, it is not computationally efficient.

However, the most significant disadvantage of these classical methods is still its weakness in preserving edges and peaks.

We use a machine learning algorithm for numerical smoothing by nonlinear diffusion equation which uses the Legendre polynomials as the kernel of LS-SVR for our proposed approach. So we introduce Legendre polynomials in the next section.
\subsection{Legendre polynomials}
\label{subsec:legendre}
Legendre polynomials are orthogonal polynomials with numerous applications, e.g., in approximation theory and numerical integration \cite{shen2011spectral}.
They can be generated recursively by the following relation \cite{delkhosh2019development}: 
\begin{equation}
    (n + 1)P_{n + 1}(x) = (2n + 1)xP_{n}(x) - nP_{n - 1}(x),\quad n \geq 1,
\end{equation}
starting with
\begin{equation}
    P_{0}(x) = 1,\quad P_{1}(x) = x.
\end{equation}
The next few Legendre polynomials are
\[
\begin{split}
    P_{2}(x) = \frac{1}{2} (3x^2 - 1),\quad P_{3}(x) = \frac{1}{2} (5x^3 - 3x) ,\quad \ldots .
\end{split}
\]

When used as a kernel of LS-SVR , the system of linear equations of the LS-SVR has many properties including sparsity, low computational cost, and some others, due to the orthogonality of this kernel. 

The Legendre polynomials are orthogonal over $[-1,1]$ with weight function $\omega(x) = 1$ as: 

\begin{equation}
    \int_{-1}^{1} P_n (x) P_m (x) \omega(x) dx = \frac{2}{2n + 1} \delta_{m n} , 
\end{equation}
where $\delta_{m n}$ is the Kronecker delta. 

The symmetry contributes to the computational cost,
\begin{equation}
\label{eq:sym}
    P_n (-x) = (-1)^n P_n (x), \quad P_n(\pm 1) = (\pm 1)^n.
\end{equation}
Therefore, $P_n (x)$ is an odd function, if $n$ is odd. Furthermore, They are bounded as:
\begin{equation}
\label{eq:bounded}
     |P_n (x)| \leq 1, \quad \forall x \in [-1,1], \; n \geq 0,
\end{equation}

which avoids error propagation.

These polynomials are the solution to Strum-Liouville differential equation as:
\begin{equation}
    ((1 - x^2)y^{\prime}_n (x))^{\prime} + n(n+1)y_n(x) = 0,
\end{equation}
with the solution $y_n(x) = P_n(x)$.

The diffusion equation for smoothing is considered on the spatial domain $[0,1]$, so we have used the shifted Legendre polynomials over $[0,1]$, given as:
\begin{equation}
    \phi_n (x) = P_n (2x-1)
\end{equation}

In this paper, we present a numerical technique for signal smoothing based on LS-SVR. We use the Legendre orthogonal polynomials as the kernel to reduce the computational complexity. To do so, first, a discretization in time is carried out by implementing the Crank-Nicolson scheme. Then, the solution of the resulting boundary value problem is approximated in the LS-SVR framework with two different approaches. The collocation points are used as the training data.

The rest of the paper is organized as follows. In section \ref{sec:svm}, we introduce support vector machines, and  least squares support vector regression (LS-SVR). Section \ref{sec:methodology} is devoted to the main methodology and the explanation of solving anisotropic diffusion equation with LS-SVR. Some numerical results are reported in Section \ref{sec:parameter-estimation}. The paper ends with some concluding remarks in Section \ref{sec:conclusion}.
\section{Support vector machines}
\label{sec:svm}
Developing algorithms that lead to learning has long been an interdisciplinary topic of research. It is proven that these algorithms can overcome many of the weaknesses posed due to the deficiency of classic mathematical and statistical approaches \cite{bishop2006pattern}.

Artificial Neural Networks (ANN) is known as one of the premier learning approaches developed in the 1940s rest on the biological neuron system of human brains. It found its widespread application almost four decades after the invention of this method in pattern recognition ever since, due mainly to its capability in extracting complex and non-linear relationships between features of different systems. Though, over time it was determined that the ANN could only perform accurately when a considerable amount of data is available for training. It lacks generalization capacity in many instances, and often a local optimal solution is given rather than a global most suitable result \cite{duda2012pattern}.

A new machine learning technique named support vector machines (SVM) introduced by Vapnik \cite{vapnik2013nature} in the early 1990s as a non-linear solution for classification and regression problems. There are some reasons behind the authority of the SVM in providing reliable outcomes. Main, its applicability is a subtle issue that accounts for the name “support vector.” Not all training examples are correspondingly relevant. Since the decision boundary only depends on the training examples nearest to it, it suffices to define the underlying model of SVM in terms of only those training examples. Those examples are called support vectors \cite{raghavan2016cognitive}. It means that unlike the ANN, this method can give accurate results with a limited number of available data. Also, unlike the ANN, the probability of having a local optimal during the training process is highly implausible when using the SVM due to having a quadratic programming problem expressed in their mathematical model. Other, its robustness upon the error of models. Unlike the ANN where the sum of squares error approach is used to decrease the errors posed by outliers, the SVM considers the trade-off parameter $c$ and insensitivity parameter $\varepsilon$ in the minimization problem which is introduced in Section \ref{subsec:svr} in order to control the error of the classification or the regression problem. Hence, the SVM can overthrow outliers by choosing the proper values of depicted parameters in the minimization problem. Finally, its computational performance and memory efficiency compared to the ANN \cite{wang2005support}.

A support vector machine seeks a precise hyperplane in a high dimensional space that can be used for either classification or regression. Support vector machines are supervised learning models used for linear and non-linear classification or regression analysis. The non-linear problem is performed using the kernel-based function for mapping input into high dimensional feature space \cite{cristianini2000introduction}.

Their application can be found in pattern recognition \cite{byun2002applications}, text classification \cite{tong2001support}, anomaly detection \cite{he2010rare}, computational biology \cite{scholkopf2004support}, speech recognition \cite{ganapathiraju2004applications}, adaptive NARMA-L2 controller for nonlinear systems \cite{kemal2021neural}, and several other applications \cite{osuna1998support}. 

Least squares support vector machines (LS-SVM) are least-squares versions of SVM, were proposed by Suykens and Vandewalle \cite{suykens1999least}.
In recent years LS-SVM regression has been used for solving ordinary differential equations \cite{mehrkanoon2012approximate} and partial differential equations \cite{mehrkanoon2015learning} which is a notable approach in approximating differential equations which. Also, in recent studies, it has been used to solve specific integral equations \cite{jani2020svr}.

\subsection{Support vector regression}
\label{subsec:svr}
The support vector regression (SVR)  possesses all the fundamentals that specify maximum margin algorithm.
It is posed to look for and optimize the generalization bounds given for regression. 

For better illustration, we first discuss the linear SVR.

\vspace{5 mm}
\begin{figure}[H]
\centering
\includegraphics[width=0.8\linewidth]{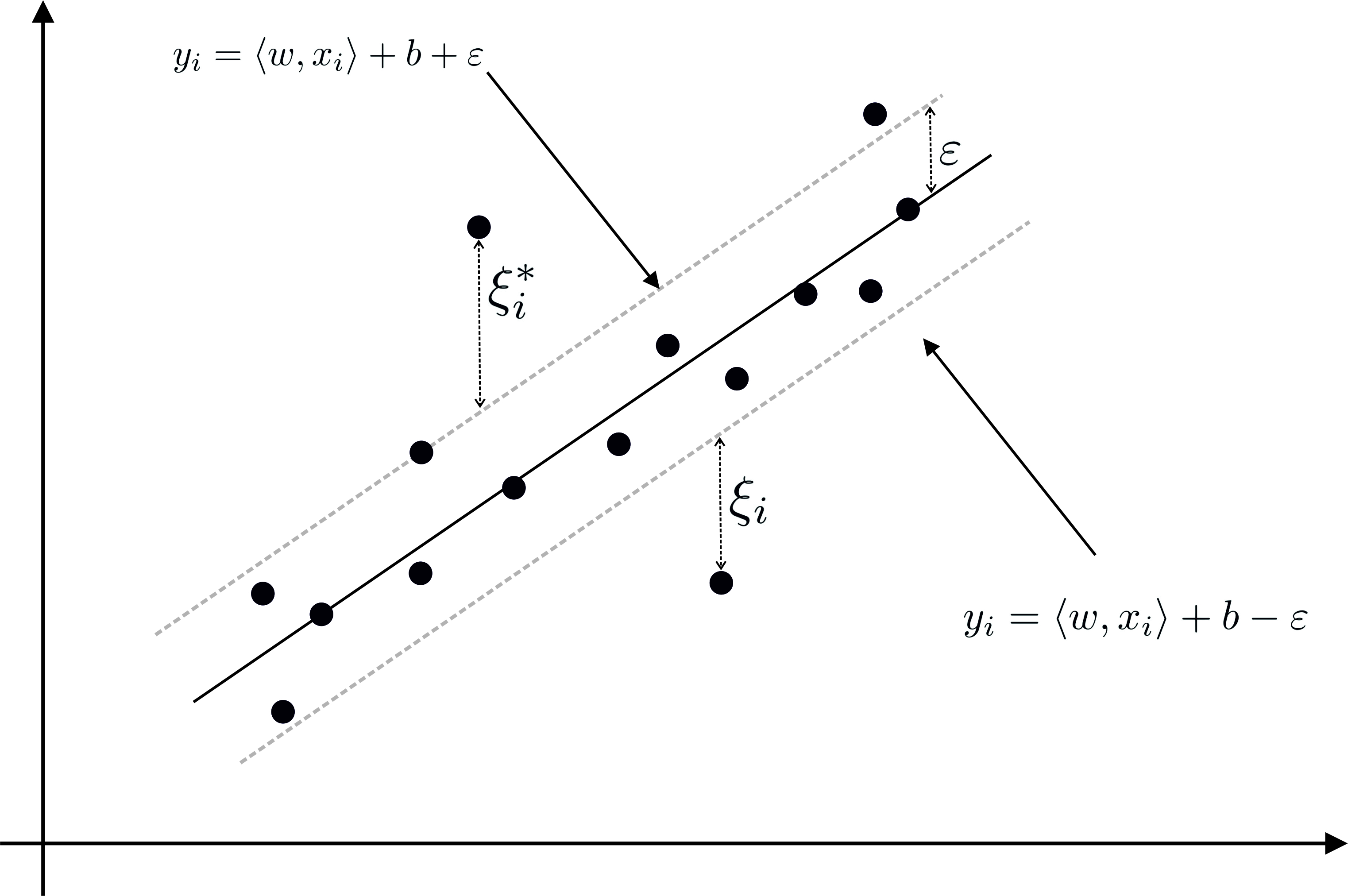}
\end{figure}
\vspace{5 mm}
The following minimization problem is solved to get the separating hyperplane:
\begin{equation*}
    \min_{w,\xi, \xi^*} \frac{1}{2}\|w\|^2 + c \sum_{i = 1}^{n}\left(\xi_i + \xi_i^* \right)
\end{equation*}

\[
    \textrm{s.t.} \begin{cases}
               y_i - \langle w, x_i \rangle - b \leq \varepsilon + \xi_i^*,\\
               \langle w, x_i \rangle + b - y_i \leq \varepsilon + \xi_i, \\
               \xi_i , \xi_i^* \geq 0, i =1,\ldots,n
            \end{cases}
\]

where $x_i$ are the training data and $y_i$ are the target values.

The solution for nonlinear SVR is to use a polynomial kernel function to transform data into a higher dimensional feature space for the purpose of making it possible to behave the same as linear SVR. In this situation, it is just needed to use a  polynomial $P(x_i)$ instead of the identity function  $x_i$.
So the minimization problem is introduced as below:
\begin{equation*}
    \min_{w,\xi, \xi^*} \; \frac{1}{2}\|w\|^2 + c \sum_{i = 1}^{n}\left(\xi_i + \xi_i^* \right)
\end{equation*}

\[
    \textrm{s.t.} \begin{cases}
               y_i - \langle w, \phi (x_i) \rangle - b \leq \varepsilon + \xi_i^*,\\
               \langle w,\phi (x_i) \rangle + b - y_i \leq \varepsilon + \xi_i, \\
               \xi_i , \xi_i^* \geq 0, i =1,\ldots,n
            \end{cases}
\]
where $\phi$ refers to the polynomial kernel function.

SVR operates linear regression in high-dimension feature space using  $\varepsilon -$intensive loss and simultaneously attempts to cut down the model complexity by minimizing $\| w \|^2$.
Slack variables $(\xi_i, \xi_i^*)$ are also added to the above constraints to measure the deviation of training samples outside $\varepsilon -$insensitive region.

It is clear that SVR estimation accuracy is determined by meta-parameters such as $c$ and $\varepsilon$, which will be tuned experimentally.

\subsection{Least squares support vector regression}
\label{subsec:ls-svr}
The main point of least squares support vector machine (LS-SVM) is to apply minimization of squared errors to the objective function in a SVM framework. It is used for either classification or regression. As mentioned earlier, in this case, we use the LS-SVM regression (LS-SVR), which solves a set of linear equations instead of solving a quadratic programming problem like the Vapnik's model.

The regression model for a dataset $(x_i,y_i), \, i = 1,\ldots,n$ is obtained by solving the following minimization problem:
\begin{equation*}
    \min_{w,e} \; \frac{1}{2}\|w\|^2 + \gamma \frac{1}{2} \|e\|^{2}
\end{equation*}

\[
    \textrm{s.t.} \;
               \langle w, x_i \rangle + b - y_i = \varepsilon + e_i,  \; i = 1,\ldots,n
\]

where $w = [w_1,\ldots,w_n]^T$, $e = [e_1,\ldots,e_n]^T$, $x_i$ and $y_i$ as the training data and target values, respectively.

The solution for nonlinear LS-SVR is to use a polynomial kernel to transform data into a higher dimensional feature space for the purpose of making it possible to behave the same as linear LS-SVR. In this situation, it is just needed to substitute the polynomial function $P(x_i)$ instead of the identity function $x_i$. So the minimization problem is introduced as below:
\begin{equation*}
    \min_{w,e} \; \frac{1}{2}\|w\|^2 + \gamma \frac{1}{2} \|e\|^{2}
\end{equation*}

\[
    \textrm{s.t.} \;
               \langle w, \phi_i (x) \rangle + b - y_i = \varepsilon + e_i,  \; i = 1,\ldots,n
\]

where $\phi$ refers to the polynomial kernel function.

\section{Methodology}
\label{sec:methodology}
In this section, we advance our approach for solving an anisotropic diffusion using LS-SVR and generating sample data to test our approach for smoothing signals. Also, the related results are mentioned in Section \ref{sec:conclusion}.

\subsection{Crank-Nicolson method} 
\label{subsec:crank-nicolson}
The Crank-Nicolson scheme is used in the finite difference method for solving the heat equation and similar PDEs like anisotropic diffusion equation to avoid the instability in the numerical results.
This method is based on the trapezoidal rule, offering second-order convergence in time. It is clear that we use this method to reduce the computational cost.
Suppose the PDE is written as:
\begin{equation}
    \frac{\partial u}{\partial t} = F(u,x,t,\frac{\partial u}{\partial x},\frac{\partial^2 u}{\partial x^2}) .
\end{equation}

Considering $u(i\Delta x,n\Delta t) = u_i^n$ and $F_i^n = F|_{(x_i,t_n)}$, the Crank-Nicolson scheme would be as follows:  
\begin{equation}
\label{eq:cn}
    \frac{u_i^{n+1} - u_i^{n}}{\Delta t} = \frac{\theta}{2} F_i^{n + 1}(u,x,t,\frac{\partial u}{\partial x},\frac{\partial^2 u}{\partial x^2}) + (1 - \theta) F_i^{n}(u,x,t,\frac{\partial u}{\partial x},\frac{\partial^2 u}{\partial x^2}),
\end{equation}
in our presentation we use $\theta = \frac{1}{2}$. 
Starting with $u^0$ as given by the initial condition, at each time step the only unknown function is $u^{n + 1}$ that is obtained by the method described below.

For the purpose of reducing the computational cost, we have implemented the Crank-Nicolson time discretization for reducing the dimension.
\subsection{Solving anisotropic diffusion equation with LS-SVR} 
\label{subsec:solve-pde}
In order to solve this PDE approximately with LS-SVR we should expand the Legendre series of the next value of time in $u$, as it follows:
\begin{equation}
\label{eq:series}
    u^{n + 1} = \sum_{i = 0}^{N} w_i \phi_i (x) .
\end{equation}
Whereas the main function $u$ depends on both $x$ values and time parameter $(t)$, it should be written as a function of two parameters, instead we use Crank-Nicolson method to discretize the time values in order to consider it as a function of one parameter $(x)$, so the Legendre series of it would be what we have written above.

By doing some manipulation, the Equation \eqref{eq:cn} can be written as:
\begin{equation}
\label{eq:cn2}
    u^{n+1} - \frac{\Delta t }{2} \left( F^{n + 1}(u,x,t,\frac{\partial u}{\partial x},\frac{\partial^2 u}{\partial x^2})\right) = u^{n} + \frac{\Delta t }{2} \left(F_i^{n + 1}(u,x,t,\frac{\partial u}{\partial x},\frac{\partial^2 u}{\partial x^2}) \right) .
\end{equation}

To save the writing we name the left side of the above equation \eqref{eq:cn2} as $L{u^{n+1}}(x)$ that is unknown and we should find it approximately later and the left side of it as $f^{n}(x)$ which is defined.

The following minimization problem must be solved in order to solve the anisotropic diffusion equation by using the LS-SVR:
\begin{equation*}
    \min_{w,e} \; \frac{1}{2}\|w\|^2 + \gamma \frac{1}{2} \|e\|^{2}
\end{equation*}

\[
    \textrm{s.t.} \;
            L{u^{n+1}}(x_i) = f^{n}(x_i) + e_{i}  ,  \; i = 1,\ldots,n
\]

where $x_i$ are the training points that are selected over $[0,1]$.
We call the above method, "Collocation LS-SVR".

An alternative idea is to apply the orthogonality of the kernel to change the minimization problem as it follows:
\begin{equation*}
    \min_{w,e} \; \frac{1}{2}\|w\|^2 + \gamma \frac{1}{2} \|e\|^{2}
\end{equation*}

\[
    \textrm{s.t.} \;
             \langle L{u^{n+1}}(x), \phi_i (x) \rangle =  \langle f^{n}(x), \phi_i (x) \rangle ,  \; i = 1,\ldots,n
\]
where $x_i$ are the training points which is selected in   $\Omega = [0,1]$ and $\langle f(x) , g(x) \rangle = \int_{\Omega} f(x) g(x) dx $.
We call the above method, "Galerkin LS-SVR".

\begin{remark}
The orthogonality of the kernel gives $\langle u, u \rangle = \sum_{i = 0}^{N} w_i^2$, so the energy of the signal in the objective function, computed by the weights, facilitates the smoothing, i.e., the term $\|w\|^2$ in minimization problem of Collocation LS-SVR or Galerkin LS-SVR that is related to Thikonov regularization plays an important role in signal smoothing.
\end{remark}

\begin{remark}
The boundedness property \eqref{eq:bounded} provides a control in error propagation and the symmetric property \eqref{eq:sym} reduces the computational cost in the method of LS-SVR.
\end{remark}

The training points may be chosen either by:
\begin{enumerate}
    \item[$\bullet$] Selecting equidistance points or,
    \item[$\bullet$] Selecting these point as the roots of Legendre polynomials.
\end{enumerate}

A good choice may be done by a meta-heuristic algorithm to take into account the oscillations in the current state of the signal.

Both minimization problems of \emph{Collocation LS-SVR} and \emph{Galerkin LS-SVR} can be reduced to a system of algebraic equations by using Lagrange multipliers method \cite{belytschko1994element} that is solved by a nonlinear solver such as the Newton method. An alternative approach is to use nonlinear optimization methods like line-search or trust-region algorithms.

We solve the anisotropic diffusion equation by using nonlinear LS-SVR with Legendre polynomial function as its kernel (\ref{subsec:ls-svr}) with particular boundary conditions as its constraints, expressed as:
\begin{itemize}
    \item[$\bullet$] $u^{n+1}(0) = u^{n}(0), \quad (n = 0)$
    \item[$\bullet$] $u^{n+1}(1) = u^{n}(1), \quad (n = 0)$
    \item[$\bullet$] $u^0 (x) = f(x), $\quad (input (noisy) signal)
\end{itemize}

By the evolution of a noisy signal in this process, it will lead to a smoothed signal.

\subsection{Overall algorithm}
\label{subsec:algorithm}
To sum up the proposed approach, the following algorithm is presented:

\begin{algorithm}[H]
\SetAlgoLined

$f(x) \leftarrow $the initial signal\;
$u^0 \leftarrow f(x)$\;
$d u^0 \leftarrow$ first derivative of $u^0$\;
$d d u^0 \leftarrow$ second derivative of $u^0$\;
$c \leftarrow$ Perona-Malik coefficient \eqref{eq:c1} or \eqref{eq:c2}\;
$\Delta t \leftarrow$ time step value (\ref{subsec:crank-nicolson})\;
\For{$n\leftarrow 0$ \KwTo $M$}{
$u^{n+1}  \leftarrow \sum_{i = 0}^{N} w_i \phi_i (x)$ \eqref{eq:series}\;

$d u^{n+1} \leftarrow$ first derivative of $u^1$\;

$d d u^{n+1} \leftarrow$ second derivative
 of $u^1$\;
 
 \For{$i\leftarrow 0$ \KwTo $N$}{
 $e_i \leftarrow L{u^{n+1}}(x_i) - f^{n}(x_i)$ \ref{subsec:solve-pde}\;
 }
 
\emph{Solving the specific minimization problem as described in the Section \ref{subsec:solve-pde}}\;
}
 
 \caption{pseudo code}
 \label{algo:steiner}
\end{algorithm}

\section{Numerical results and discussion}
\label{sec:parameter-estimation}
In this section, two illustrative numerical examples will be provided to clarify the efficiency of the proposed approach. First, we introduce the data generation method, which is used for our computational experiments. Next, these examples are reported in the rest of the paper in detail.

\subsection{Data generation} 
\label{subsec:data-generation}
Normally, we can simulate a noisy signal by adding white noise to the spectral peak which can be defined as Gaussian peaks, Lorentzian peaks, and their combinations. 

Due to the fact that Pearson VII function is equivalent to another parameterization of the Student-t distribution function, it can generate the aforementioned peaks as well \cite{li2018spatial} which is defined as below:

\begin{equation}
    g(x) = \frac{A}{[1 + 4 \left( \frac{x - \mu}{\sigma} \right)^2 \left( 2^{\frac{1}{q}} - 1 \right)]}
\end{equation}

where $A$, $\mu$, and $\sigma$ are related to the height, position, and width of the peak, respectively.
As a rule, when the parameter $q$ is equal to $1$, the function is a Lorentzian peak. As the value of parameter $q$ increases, the function converts to the Gaussian peak.

As we mentioned above, we need to add white noise to these peaks to generate the required data, which the white noise can be created by generating random numbers in a particular range by "GenerateGaussian" function in Maple and using cubic spline to connect these points to get a continuous function.

\subsection{Numerical observations}
\label{subsec:results}
Here, we reported the numerical results by implementing our approach using Maple 2018 on a laptop with configuration: Intel(R) core(TM) i7-4720HQ CPU @ 2.60 GHz and 12GB of RAM.

As a convention in all test cases, the red plot is the noisy signal and the blue one is the smoothed signal.

For the first test case, we have used the data generation method we discussed in the previous section (\ref{subsec:data-generation}). With the particular parameters as expressed here: 
\begin{equation}
    A = 2, \qquad q = \frac{1}{5}, \qquad \sigma = \frac{1}{2}, \qquad \mu = \frac{1}{2}
\end{equation}

By the evolution of the first test case generated data as an input signal in the algorithm \ref{algo:steiner} it will lead to smoothed signal as it is noticeable in the following plots:
\vspace{5 mm}
\begin{figure}[H]
     \centering
     \begin{subfigure}[b]{0.495\textwidth}
         \centering
         \includegraphics[width=\textwidth]{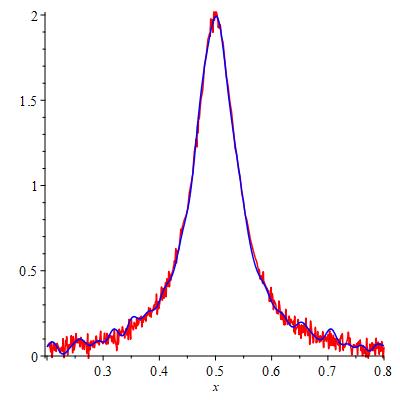}
         \caption{After 0.1 seconds, SNR = 14.49373.}
         \vspace{5mm}
     \end{subfigure}
     \hfill
     \begin{subfigure}[b]{0.495\textwidth}
         \centering
         \includegraphics[width=\textwidth]{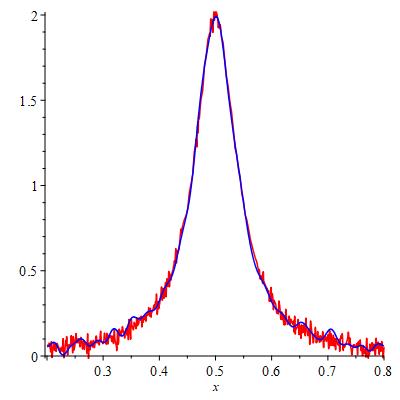}
         \caption{After 0.2 seconds, SNR = 16.68975.}
         \vspace{5mm}
     \end{subfigure}
\vspace{5mm}
     \begin{subfigure}[b]{0.495\textwidth}
         \centering
         \includegraphics[width=\textwidth]{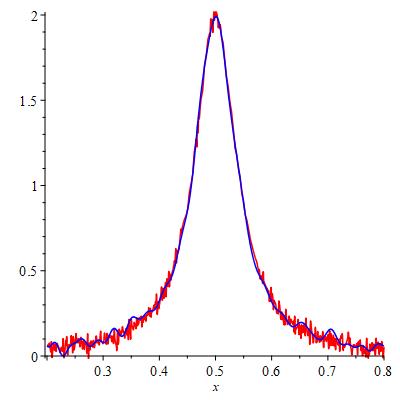}
         \caption{After 0.3 seconds, SNR = 22.23075.}
     \end{subfigure}
     \hfill
     \begin{subfigure}[b]{0.495\textwidth}
         \centering
         \includegraphics[width=\textwidth]{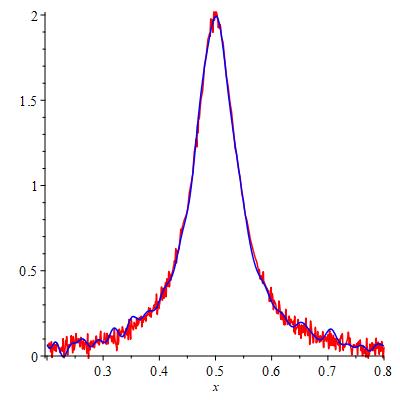}
         \caption{After 0.4 seconds, SNR = 26.67823.}
     \end{subfigure}
        \caption{Displaying the evolution of the first test case over $[0.2,0.8]$, the elapsed evolution time and calculated SNR are labeled below of each plot.}
        \label{fig:first}
\end{figure}
\vspace{5 mm}

As can be perceived from the SNR values of the first test case shown in Figure \ref{fig:first}, the input signal has been smoothed well.

Furthermore, to prove the strength of our presented approach, we have made a comparison between our proposed method and the Savitzky-Golay method here. We have done the Savitzky-Golay quadratic smoothing method with the filter width of 9; it can be noted that this size of filter width resembles more effective than other sizes, empirically.

For this purpose, the following plot and report are provided:
\vspace{5 mm}
\begin{figure}[H]
\centering
\includegraphics[width=0.5\textwidth]{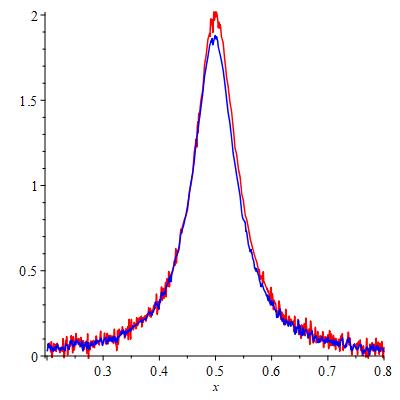}
\caption{Savitzky-Golay quadratic smoothing for the first test case with the filter width of 9, SNR = -2.40262.}
\end{figure}
\vspace{5 mm}

As it is obvious, our proposed method has smoothed the noisy signal more efficient than the Savitzky-Golay method without demolishing the peak, while the Savitzky-Golay method has destroyed the peak without omitting all unwanted features well, according to the SNR values.

Note that, since the classical signal smoothing methods, which are based on moving average concept same like the Savitzky-Golay method, destroys the beginning and the end of the signal, and because usually, the critical data does not exist at the beginning and the end of a signal, we have reported the information over $[0.2,0.8]$, in order to gain a fair comparison.

For the second test case, we have done the same method as the first test case in order to generate data by modifying some parameters, as stated here: 
\begin{equation}
    A = 2, \qquad q = 2, \qquad \sigma = 1.5, \qquad \mu = 3
\end{equation}

By the evolution of the second test case generated data as an input signal in the algorithm \ref{algo:steiner} it will lead to smoothed signal as it is noticeable in the following plots:
\vspace{5 mm}
\begin{figure}[H]
     \centering
     \begin{subfigure}[b]{0.495\textwidth}
         \centering
         \includegraphics[width=\textwidth]{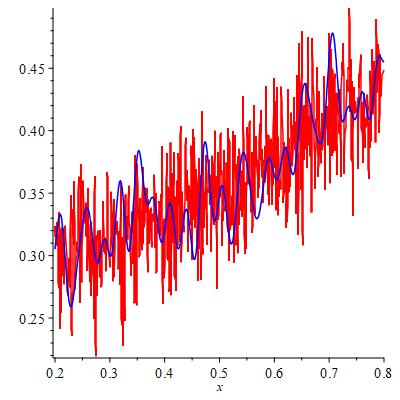}
         \caption{After 0.1 seconds, SNR = 29.58323.}
         \vspace{5mm}
     \end{subfigure}
     \hfill
     \begin{subfigure}[b]{0.495\textwidth}
         \centering
         \includegraphics[width=\textwidth]{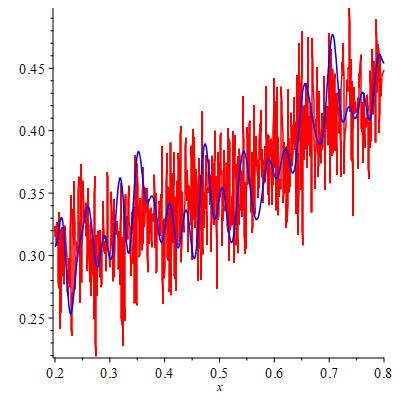}
         \caption{After 0.2 seconds, SNR = 31.53024.}
         \vspace{5mm}
     \end{subfigure}
\vspace{5mm}
     \begin{subfigure}[b]{0.495\textwidth}
         \centering
         \includegraphics[width=\textwidth]{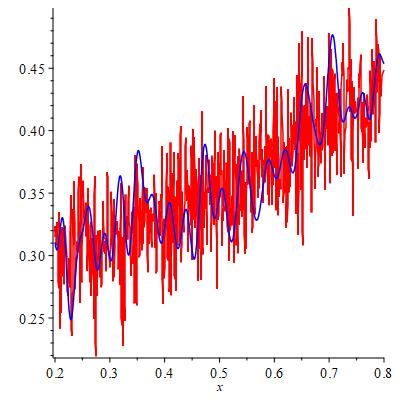}
         \caption{After 0.3 seconds, SNR = 36.64582.}
     \end{subfigure}
     \hfill
     \begin{subfigure}[b]{0.495\textwidth}
         \centering
         \includegraphics[width=\textwidth]{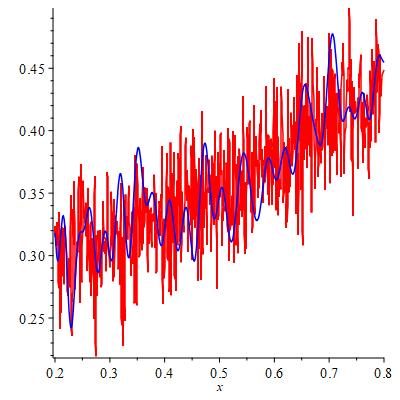}
         \caption{After 0.4 seconds, SNR = 40.18162.}
     \end{subfigure}
        \caption{Displaying the evolution of the second test case over $[0.2,0.8]$, the elapsed evolution time and calculated SNR are labeled below of each plot.}
        \label{fig:second}
\end{figure}
\vspace{5 mm}

As can be perceived from the SNR values of the second test case shown in Figure \ref{fig:second}, the input signal has been smoothed well.

As in the first test case, we made a comparison with Savitzky-Golay method, for the second test case, we have the following plot and report to address a comparison with our proposed method.

\vspace{5 mm}
\begin{figure}[H]
\centering
\includegraphics[width=0.5\textwidth]{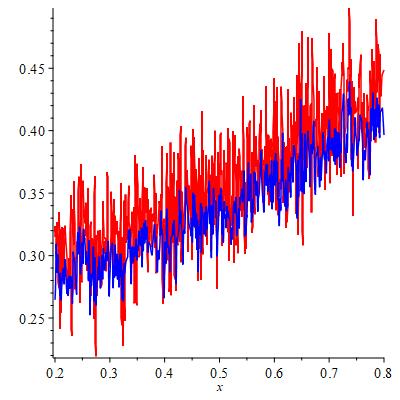}
\caption{Savitzky-Golay quadratic smoothing for the second test case with the filter width of 9, SNR = 14.36374.}
\end{figure}
\vspace{5 mm}

As reported from the SNR values show in the second test case, our proposed method performs better than the Savitzky-Golay method in the signal smoothing process.

According to numerical results, it is clear that the signal smoothing is done successfully by the presented approach:
\vspace{5mm}
\begin{table}[H]
\centering
\begin{tabular}{@{}c|cc@{}}
                                & \textbf{the 1st test case} & \textbf{the 2nd test case}  \\ \hline
    \textbf{after 0.1 seconds}          & 14.49373          & 29.58323       \\ \hline
    \textbf{after 0.2 seconds}          & 16.68975          & 31.53024       \\ \hline
    \textbf{after 0.3 seconds}          & 22.23075          & 36.64582      \\ \hline
    \textbf{after 0.4 seconds}          & 26.67823          & 40.18162      \\
            \end{tabular}
            \vspace{5mm}
        \caption{The SNR values of test cases over time elapsed}
        \label{tab:snr-values}
\end{table}
Consequently, as time goes by, it becomes smoother as the SNR value increases. In other words, higher SNR value in a test case means it has less noise.

One of the advantages of this method is that we can smooth the signal as far as we need by altering the hyperparameters. It is essential to mention that the cease of the elapsed time of the input signal evolution to become smooth, should be done by monitoring the SNR value.

\section{conclusion}
\label{sec:conclusion}
The proposed approach for signal smoothing based on anisotropic diffusion is presented in a machine learning framework by using the LS-SVR with orthogonal kernel is proved to be highly competent for signal smoothing. In order to optimize our method's computational cost, we have used the Crank-Nicolson method for discretization in time besides using orthogonal polynomials as the LS-SVR kernel. It is clear that our approach is based on evolution in time; this is important because, in this method, the smoothing rate can be controlled. Numerical experiments confirm the validity and efficiency of our presented approach. It is seen that the method does not destroy edges rather preserving it. We can consider it as a capable algorithm and could be used as an adaptive signal smoothing method.

%\begin{acknowledgements}
%\end{acknowledgements}

 \section*{Conflict of interest}
 The authors declare that they have no conflict of interest.

\bibliographystyle{spmpsci}      % mathematics and physical sciences
\bibliography{main}   % name your BibTeX data base

%\begin{thebibliography}{}
% etc
%\end{thebibliography}

\end{document}